\documentclass[3p,onecolumn]{elsarticle}
\usepackage{lineno,hyperref}
\usepackage{threeparttable}
\modulolinenumbers[5]
\journal{Journal of Magnetism and Magnetic Materials}
\usepackage{times}
\usepackage{color}

\newcommand{\Heb}{H_{eb}}
\newcommand{\Hc}{H_{c}}

\newcommand{\Jfm}{J_{FM}}

\newcommand{\Mshif}{M_{AF,int}}
\newcommand{\Rc}{R_c}
\newcommand{\Lc}{L_c}
\newcommand{\tsh}{t_{sh}}

\begin{document}
\begin{frontmatter}
	\title{Magnetic properties of nanowires \\ with ferromagnetic core and antiferromagnetic shell}

	\author[address1]{A.~Patsopoulos\corref{corradd} }
	\ead{apats@phys.uoa.gr}

	\author[address2]{D.~Kechrakos}
    \ead{dkehrakos@aspete.gr}

	\author[address2]{N.~Moutis}

	\cortext[corradd]{Corresponding author}

    \address[address1] {Department of Physics, National and Kapodistrian University of Athens, GR-157 84 Athens, Greece} 

	\address[address2]{Department of Education, School of Pedagogical and Technological Education (ASPETE), Athens, GR-14131}
\begin{abstract}
We present a Monte Carlo study of the magnetic properties of thin cylindrical nanowires composed of a ferromagnetic core and an antiferromagnetic shell implementing a classical spin Hamiltonian. We address systematically the impact of interface exchange coupling  on the loop characteristics and the magnetization reversal mechanism.
We study the effect of shell polycrystallinity on the characteristic fields of the isothermal hysteresis loop (coercivity, exchange-bias). 
We demonstrate that coupling to a polycrystalline antiferromagnetic shell  increases the critical core  diameter  for transition from transverse to vortex domain walls.
\end{abstract}
\begin{keyword}
	exchange bias \sep 
	core-shell nanowires\sep 
	domain wall\sep 
	Monte Carlo.
	\PACS 75.60.Jk \sep 75.75.Jn \sep 75.75.Fk \sep 75.78.Fg  
\end{keyword}
\end{frontmatter}
\section*{Highlights}
\begin{description}
	\item[$\bullet$] Exchange bias in Co/CoO core-shell nanowires 
	\item[$\bullet$] Shell granularity and magnetization reversal 
	\item[$\bullet$] Critical diameter for transverse-to-vortex  wall transition
\end{description}

\section{Introduction}
Magnetic nanowires have potentials for a variety of innovative technological applications\cite{sko03}, ranging from magnetic recording\cite{sel01} to biomedicine\cite{bau04,tar06}, owing to their enhanced magnetic anisotropy that stems from their elongated shape. 
The controllable one-dimensional character of domain wall propagation in these nanosystems finds application in racetrack memory technology \cite{par08}.
Also, the quasi one-dimensional shape leads to a complex reversal mechanism, consisted of domain wall (DW) nucleation, propagation and annihilation\cite{fer99,sel01,thi06}. 
In addition, the diameter of a FM nanowire was shown \cite{hin00a,wie04} to control the domain wall character. 
In particular, a transition from a transverse domain wall (TDW) to a vortex domain wall (VDW) was predicted as the diameter of the wire increases beyond the exchange length, with direct implications on the domain wall mobility\cite{wie04}. 
	Also, recent experiments on cylindrical cobalt nanowires and supporting micromagnetic simulations demonstrated the transition from uniform to vortex remanent state with increasing nanowire diameter\cite{iva16}, underlying the possibility of tailoring the magnetization state through the geometrical characteristics of the nanowire.

Tailoring the magnetic anisotropy remains a central issue  in studies of nanowires. Periodic chemical modulation\cite{ber17} and periodic morphological modulation\cite{pal15,fer18} are among the recently reported methods. 
On the other hand, the exchange bias effect\cite{mei56,mei57} has long been recognized as a means to tailor the hysteresis characteristics of nanostructured magnetic materials\cite{nog05,igl08}. 
Recent experiments, demonstrated exchange bias behavior in Py nanowires\cite{buc15} and oxidized Co nanowires \cite{mau09,tri10,gan17} and nanotubes\cite{pro13}, with the characteristic accompanying effects (loop shift, training effect, etc.).
In our previous works\cite{pat17,pat18}, we studied numerically the modifications of the magnetic properties introduced by the interface exchange coupling of a FM core to an AF shell in cylindrical Co/CoO nanowires. 
We showed that a monocrystalline CoO shell acts as a sequence of nucleation centers introducing a secondary mechanism for magnetization reversal which acts in synergy with the transverse domain wall propagation promoting the reversal\cite{mau09,pat17}. 
Furthermore, we showed that shell granularity "softens" the AF shell causing  strong drag of the shell interface moments by the core moments, with a concomitant coercivity enhancement and exchange bias suppression\cite{pat18}. 
Finally, granularity was shown to introduce an off-axis unidirectional anisotropy leading to maximum exchange bias field in an off-axis direction\cite{pat18}. 

In the present work,  we focus on the effect of interface exchange and shell granularity on the transition between different magnetization reversal modes in core-shell nanowires and demonstrate that the presence of exchange bias modifies the critical diameter for transition from TDW to VDW.
\section{Model and Simulation Method}
\begin{figure}[htb!]
	\centering
	\includegraphics[width=0.65\linewidth]{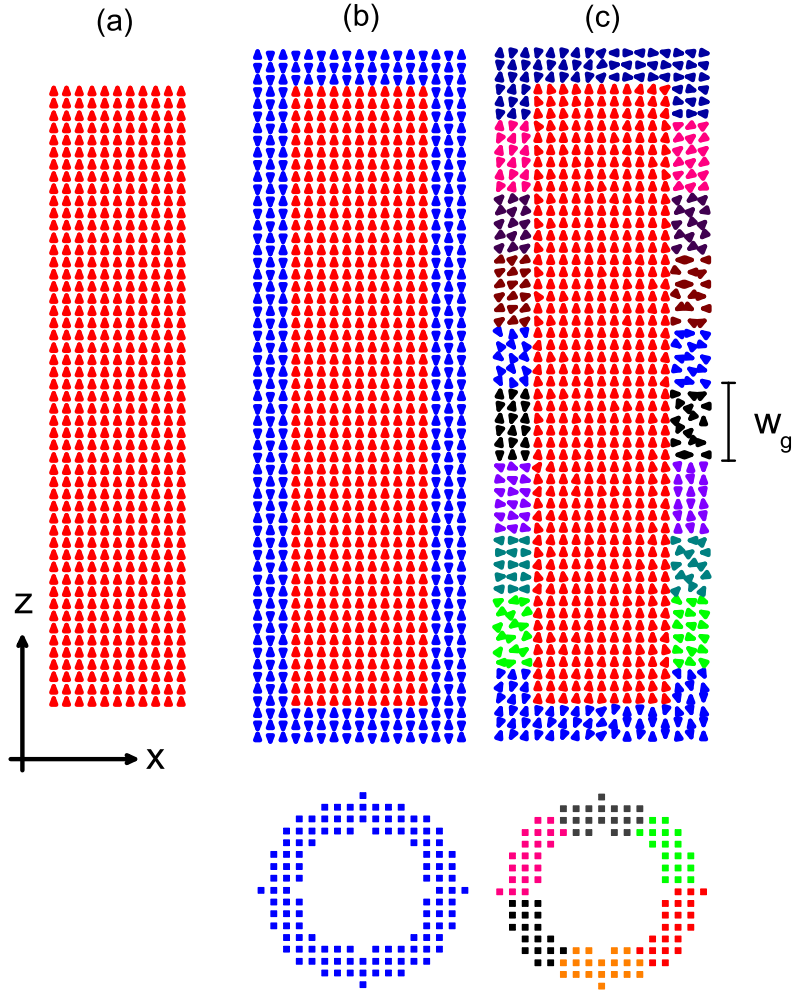}
	\caption{ 
		(Color online) Cutting planes (top) and shell cross sections (bottom) of cylindrical magnetic nanowires with $R_c=15nm$, $L_c=150nm$ and $\tsh=9nm$ at the FC state.
		(a) $FM$ nanowire,
		(b) $FM/AF$ nanowire with a monocrystalline shell and
		(c) $FM/AF$ nanowire with a polycrystalline shell containing $N_g=60~(N_z=10, N_{\phi}=6)$ grains of width $w_g=15nm$ and $18nm$.
		Notice that the shell moments freeze along the easy axis of the parent grain.
	}
	\label{fig:sites2}
\end{figure}

	We consider nanowires with cylindrical shape of radius $R$ and length $L$ along the z-axis.
An internal homoaxial cylinder with radius $R_c = R-t$ and length $L_c=L-t$ constitutes the FM core and the outer hollow cylinder with thickness $t$ the AF shell.
	We discretize the whole cylinder by means of a simple cubic grid with cell size $a$.
To model the shell polycrystallinity (granularity) we divide the shell in $N_z$ cylindrical slices along the z-axis and each slice in $N_\phi$ circular sectors (Fig.~\ref{fig:sites2}c).
	This simple scheme generates  $N_g=N_z\cdot N_\phi$ grains of almost equal size in the shell.
	On the other hand, the FM core is treated as a single crystal in all our simulations.

The magnetic structure is described by a classical Heisenberg Hamiltonian. The total energy of the magnetic system is decomposed in contributions from individual cells (sites) as $E=\sum_i E_i$, where
\begin{eqnarray}
E_i = 
-\frac{1}{2} \widehat{S}_i \cdot \sum_{<j>} J_{ij}  \widehat{S}_j 
-K_i (\widehat{S}_i \cdot \widehat{e}_i )^2  \nonumber \\
- H S_{i,z}  
-\frac{1}{2} g \widehat{S}_i \cdot \sum_{j}  \textit{\textbf{D}}_{ij} \cdot \widehat{S}_j. 
\label{eq:energy}
\end{eqnarray}
Hats indicate unit vectors and bold symbols $3\times 3$ matrices in Cartesian coordinates. 
The $1/2$ prefactors in the first and fourth terms of Eq.~\ref{eq:energy} account for the double-counting of energy contribution from pairs of sites.
The first term in Eq.~\ref{eq:energy} is the exchange energy between first nearest neighbors (1nn) sites. 
The exchange constant  $J_{ij}$ takes the values $J_{FM}$, $J_{AF}$ and $J_{int}$ depending on whether sites $i$ and $j$ belong to the $FM$, the $AF$ or the interface region, respectively. 
The second term in Eq.~\ref{eq:energy} is the uniaxial anisotropy energy.
The easy axes ($\widehat{e_i}$) of the core sites are taken along the cylinder axis.
	A random easy axis is attributed to all sites belonging to the same grain, as an  approximation to the misorientation of real grains.
The anisotropy constant $K_i$ takes the values $K_{FM}$ and $K_{AF}$ depending on the location of site $i$.    
The third term in Eq.~\ref{eq:energy} is the Zeeman energy due to the applied field along the cylinder axis and the fourth term is the dipolar energy with strength $g$.
	For computational efficiency, we decompose the local dipolar field 
	$H^{d-d}_i\equiv g \sum_{j}  \textit{\textbf{D}}_{ij} \cdot \widehat{S}_j$
	into a near-field part that extends up to 3rd nearest neighbor cells and a far-field part that is treated in a mean field approximation\cite{pat17}.
	We use micromagnetic parameters typical of  cobalt $A=1.3\cdot10^{-11}J/m, M_s=1.4\cdot10^6 A/m$, and $K_u=3\cdot10^5 J/m^3$. 
	The shell material is cobalt oxide for which we assume for simplicity, the same net moment per cell $M_{s,CoO}=M_{s,Co}$, reduced exchange stiffness $A_{CoO}=$0.5$A_{Co}$ as dictated by the relation between their critical temperatures $T_N \sim 0.5 T_C$ and strong effective anisotropy energy density $K_{u,CoO}=10 K_{u,Co}$. 
	We use a grid cell size $a=3$nm, which is smaller than the estimated exchange length of cobalt, $\lambda \approx 3.3$ nm.
	We scale all energy parameters entering Eq.~\ref{eq:energy} by $\Jfm$ and  
	the relevant parameters entering the simulation eventually read:
$J_{AF}/\Jfm$ = -0.5,~ 
$J_{int}/\Jfm$ = -0.5,~ 
$K_{FM}/\Jfm$ = 0.1,~ 
$K_{AF}/\Jfm$ = 1.0 and 
$g/\Jfm=0.07$.
We numerically simulate the field-cooling process
(Fig.~\ref{fig:sites2}) and the isothermal hysteresis loop implementing the Metropolis Monte Carlo algorithm with single spin updates, as previously described\cite{pat17}.
\section{Results and Discussion}
\subsection{Isothermal hysteresis loops}
\begin{figure} [htb!]
	\centering
	\includegraphics[width=0.65\linewidth]{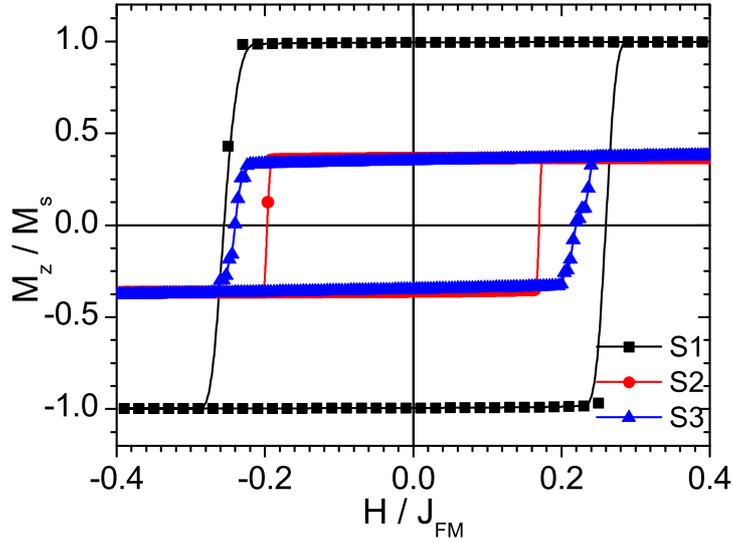}
	\caption{ 
		(Color online) Isothermal hysteresis loops of nanowires: 
		Squares (black): $FM$ nanowire.
		Circles (red): $FM/AF$ nanowire with monocrystalline shell. 
		Triangles (blue): $FM/AF$ nanowire with polycrystalline shell ($N_g=60$).  
		In all cases the $FM$ region has dimensions
			$R_{c}=15nm$ and $L_{c}=150nm$, while the shell thickness in core/shell systems is $t=9nm$.
	}
	\label{fig:Loops}
\end{figure}

First, we study the macroscopic magnetic behavior of the nanowires, by calculating the low-temperature hysteresis loops and the characteristic fields of the loop, namely the coercivity ($\Hc$) and the exchange-bias ($\Heb$). 
We compare three different types of nanowires: a FM nanowire (S1),  a core-shell nanowire with crystalline shell (S2) and a core-shell nanowire with 
	monocrystalline core and a
polycrystalline shell containing $N_g=60$ grains (S3). The size of grains in S3 is approximately equal to the DW width.
The impact of the $AF$ shell on the magnetic properties of FM nanowires, can be initially extracted by comparing the properties of the FM nanowire (S1) and the monocrystalline core/shell nanowire (S2). Finally, we gain a complete picture of the shell contribution by examining the differences emerging from the shell polycrystallinity.
In Fig.~\ref{fig:Loops} we show isothermal hysteresis loops at low temperature. 
We observe a symmetric square loop for the FM nanowire, while a clear horizontal shift of the loops is observed in core/shell samples as the outcome of the FC process. 
Core/shell systems also exhibit a drastic decrease in coercivity ($Hc$). 
This phenomenon occurs due to the existence of unsatisfied bonds in the $AF$ shell. 
More specifically, in an ideal AF shell, like the monocrystalline shell, the satisfied and unsatisfied bonds are almost equal, because differences occur only at the edges of the finite nanowire. 
Thus the core-shell nanowire (S2) is weakly uncompensated.
However, the transition probability for thermal activation of a satisfied and an unsatisfied bonds are not equal, according to  Arrhenius law ($P\sim exp(-dE/k_{B}T$).
As a result the transition probability to a new state is greater for the unsatisfied bonds, which act as nucleation centers and determine the magnetic behavior leading to the observed decrease of the coercive field relative to the FM case.

In addition, by comparison of the S2 and S3 in Table~\ref{table:hchb}, it becomes clear that granularity leads to an increase of coercivity and decrease of the exchange-bias field. 
This trend with increasing granularity was demonstrated previously \cite{pat18}, and is the outcome of two distinct physical factors, namely the response of the shell-interface magnetization ($\Mshif$) to the applied field and the actual value of $\Mshif$ at the FC state. 
A large value of $\Mshif$ indicates an increase in the number of satisfied bonds at the FC state. 
As a last remark, we observe that the loop of S1 remains wider than the loop of S3. 
In other words, the shell granularity despite the increase of coercivity that it produces, it cannot restore the value of coercivity corresponding to the bare FM sample. 
This implies that the number of unsatisfied bonds are still important in polycrystalline shell systems. 
The last result is in accordance with experimental results on isolated Co nanowires\cite{mau09}.
\begin{table}    
	\centering
	\caption{
		Results for nanowires with $\Rc=15nm$, $\Lc=150nm$ and $\tsh=9nm$.
	}
	\begin{threeparttable}
		\begin{tabular}{c c c c c r}   
			\hline
			Sample 
			&NW type   
			&$\Hc$          \tnote{(1)}  
			&$\Heb$         \tnote{(1)}  
			&$\Mshif$       \tnote{(2)}  \\
			\hline
			S1   &FM          &0.285  & -      & -     \\
            S2   &FM/AF       &0.182  &-0.014  &0.015  \\
            S3   &FM/AF-poly  &0.230  &-0.011  &0.052  \\
			\hline	
		\end{tabular}
		\begin{tablenotes}
			\begin{small}
				\item[(1)] Field values in units of $\Jfm$ 
				\item[(2)] shell-interface magnetization (per spin) at the FC state
			\end{small}
		\end{tablenotes}
	\end{threeparttable}
	\label{table:hchb}
\end{table}
	\subsection{Magnetization reversal mechanism}
We discuss next, the impact of shell polycrystallinity on the underlying magnetization reversal mechanism.
\begin{figure} [htb!]
	\centering
	\includegraphics[width=0.65\linewidth]{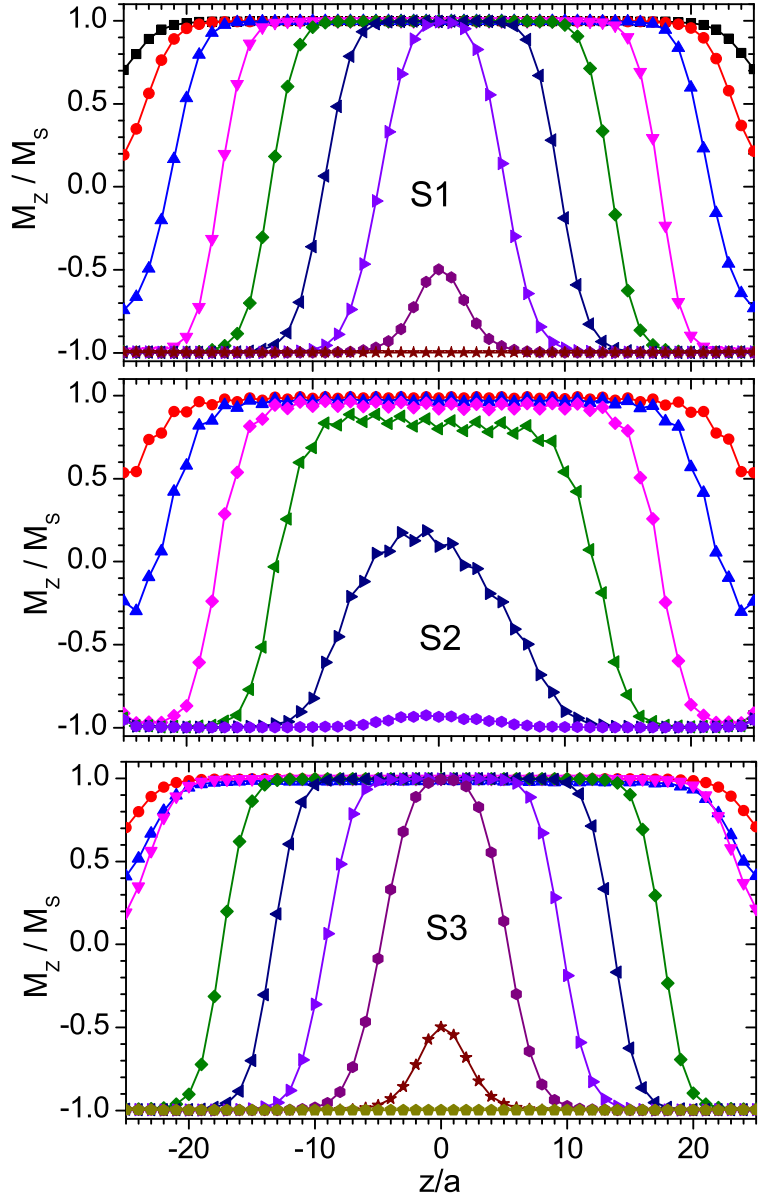}
	\caption{
		(Color online) Time-evolution of magnetization profile under application of a reverse field $H/\Jfm=-4.0$ for bare (S1) and core/shell (S2,S3) nanowires. Snapshots are taken every $\Delta$t=200 MCSS starting at $t_0$=200 MCSS.
		Structural parameters as in Table~\ref{table:hchb}.
	}
	\label{fig:Mprofs3}
\end{figure}
Magnetization reversal in FM nanowires proceeds by nucleation of a pair of domain walls at the free ends of the nanowire, their propagation with opposite velocities and eventually their merge at the center of the wire \cite{fer99,hin00a,thi02}. 
The coupling to an AF shell (S2) modifies the reversal mechanism, as previous experimental\cite{mau09} and numerical\cite{mau09,pat17} works have demonstrated.
In an ideal FM/AF nanowire (S2), as it was mentioned above, the unsatisfied bonds at the interface act as nucleation centers of a secondary magnetization reversal mechanism. 
This behavior, apart from shrinking of the hysteresis loop, can also be seen in Fig.~\ref{fig:Mprofs3}b as a lowering of the core magnetization in the central region between the two domain walls. 
This second reversal mechanism acts in synergy to the domain wall propagation and accelerates the reversal of the core magnetization.

\begin{figure}[htb!]
	\centering
	\includegraphics[width=0.65\linewidth]{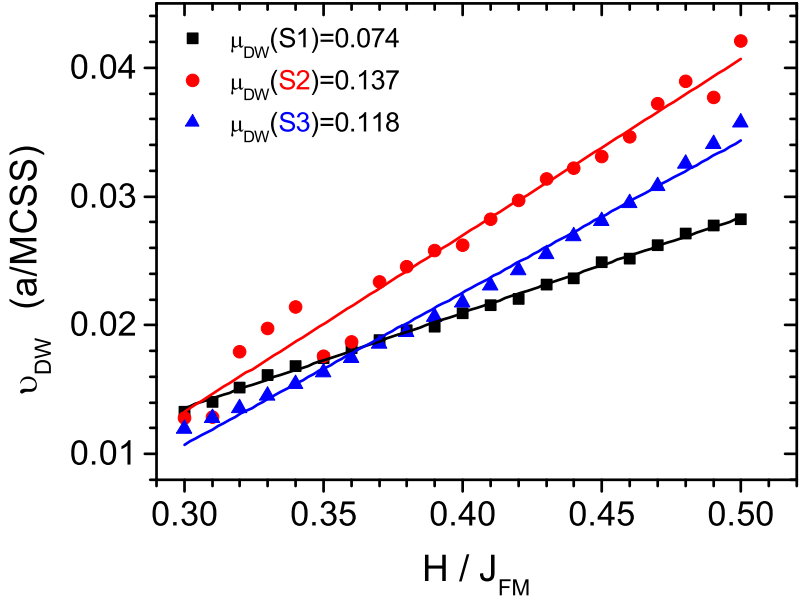}
	\caption{
		(Color online) Field-dependence of DW velocity of bare (S1) and core/shell (S2,S3) nanowires.
		Straight lines are linear fit to the data.  
		DW mobility (slope) increases due to the $AF$ shell, but  shell granularity leads to reduction of DW mobility relative to the monocrystalline shell. 
		Mobility values indicated are in $a/MCSS$. Structural parameters as in Table~\ref{table:hchb}.
	} 	
	\label{fig:veloc}
\end{figure}

As one can readily observe in Fig.~\ref{fig:Mprofs3}, this secondary mechanism is absent in the S3 nanowire, where the reversal proceeds by clear domain wall propagation. 
Suppression of the secondary mechanism is due to the effective magnetic softening of the shell magnetization in the S3 nanowire, which no longer acts as a collection of nucleation centers for magnetization reversal. 
Futhermore, on a microscopic level, the number of satisfied interface bonds is substantially higher for the S3 nanowire. 
These bonds being in their lowest energy state oppose their reversal under the applied field, acting as soft pinning centers for domain wall propagation\cite{pat18}.
Thus, drag of the AF interface moments  and increased $\Mshif$ magnitude due to polycrystallinity are the two factors acting in synergy to suppress the domain wall mobility in polycrystalline samples relative to samples with crystalline shells (Fig.~\ref{fig:veloc}). 
\subsection{Character of domain walls}
Another aspect that determines the magnetization reversal is the character of the domain walls. 
Hinzke and Nowak \cite{hin00a} showed that the ferromagnetic nanowires can support either TDW or VDW, depending on the size of their diameter. 
Thin nanowires support transverse domain walls. 
As the diameter increases the system minimizes the total energy by forming closure domains due to dipolar energy leading to a transition from T to V domain walls. 
Here we examine the impact of exhange bias on the transition from TDW to VDW in magnetic nanowires.   
We consider long enough wires ($L=300$nm) so that the coercive and exchange bias fields are almost independent of the nanowire length\cite{pat17}.

\begin{figure}
\centering \includegraphics[width=0.55\linewidth]{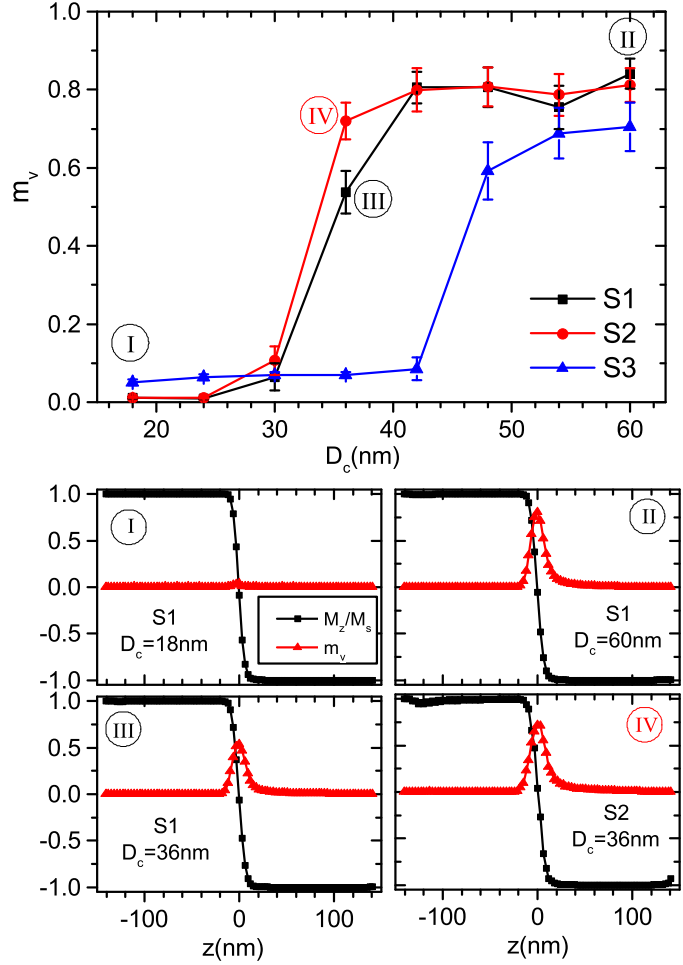}
\caption{
	(Color online) Dependence of magnetization vorticity on core diameter (upper panel) and the  corresponding profiles along the nanowire axis when the DW passes from the center of the nanowire (lower panel). The length of nanowire core is
			 $\Lc=300$nm and the shell thickness $\tsh=9$nm.
	Error bars are obtained from an ensemble of 50 different samples with (S3) or without (S1,S2) structural disorder.
	}
\label{fig:v_d}
\end{figure}

The nanowire is brought initially at the FC state and then a field with direction opposite to the cooling field is applied. 
During the relaxation of the system under application of the reverse field, we freeze the moments at one end of the nanowire, leaving only one free end. 
Under these boundary conditions, a single DW nucleates and propagates in the nanowire instead of a pair of opposite propagating walls. 
The character of the DW (Transverse or Vortex) is quantified by the parameter $m_v$ that measures the  magnetization vorticity\cite{ter16} and is defined as
\begin{equation}
m_v(z) =\sum_{i\in FM} 
[ \widehat{\rho}_i \times  \widehat{S}_i ]_z \delta(z_i-z) / 
\sum_{i\in FM}\delta(z_i-z) 
\end{equation}
where $\rho_i$ the $xy-$projection of the position vector on the $i$-th cell.
In particular, $m_v=0$ for a TDW and $m_v=\pm1$ for a VDW with the sign indicating the direction of magnetization rotation in the $xy$-plane.
We calculate the vorticity of nanowires with fixed length ($L_c=300$nm) at the instant of vanishing magnetization ($M=0$), namely when the DW is located at the center of the nanowire ($z=0$).
In Fig.~\ref{fig:v_d} we show the dependence of magnetization vorticity $m_v$ on core diameter, for the three different types of wires in order to highlight the impact of the AF shell and shell granularity on the  transition from T to V domain walls. 
	A transition in the wall character is clearly observed for all types of nanowires. 
Notice that the vorticity curve does not reach unity for diameters up to $D=60$nm, due to thermal (statistical) effects inherent to the Monte Carlo method and the finite length of the nanowires modeled. 
We notice that in core/shell systems the transition takes place in slightly smaller diameters ($D_{crit}\approx 36$nm) than in bare FM nanowires ($D_{crit}\approx 42$nm). 
We understand the formation of vortex structure at the domain wall center as the outcome of energy minimization achieved predominantly by minimization of the dipolar energy of the system, which favors the formation of in-plane vortex structures in cylindrical nanostructures.
The coupling of the FM core to an AF shell (S2) is approximately equivalent to the action of a (frozen) local bias fields on the core-interface spins which promotes the reversal of core-interface moments, as discussed in the previous section. 
Consequently, the core/shell nanowire  appears magnetic softer and favors formation of vortex walls at slightly smaller diameter.  
Shell granularity (S3) is shown to have a strong effect on the transverse-to-vortex domain wall transition. In this case, the core-interface spins drag the shell-interface spins, which renders them magnetically harder and consequently these moments prohibit formation of a vortex structure. 
\section{Summary}
We have studied the isothermal magnetic hysteresis of cylindrical Co-CoO nanowires with core/shell morphology using a micromagnetic model and Monte Carlo simulations.
We showed that the coupling to the antiferromagnetic shell leads to emergence of a weak exchange bias effect and a severe suppression of coercivity compared to the bare FM nanowire. 
However, shell polycrystallinity  produces  coercivity  values comparable to the bare FM nanowire and consequent suppression of the exchange-bias field relative to the monocrystalline shell. 
This behavior is mainly attributed to enhanced drag of the AF interface spins during reversal of the core spins. 
The same mechanism is responsible for reduction of domain wall mobility.
Finally, the coupling to perfect AF shell only weakly decreases the critical diameter for T-to-V domain wall transition, while the presence of shell polycrystallinity increases the critical diameter.
This latter finding is anticipated to have implications in the choice of nanowire diameters for tailoring the character of the DW as in diameter modulated nanowires\cite{pal15,arz17}.
	Finally, we mention additional microstructural factors, such as core polycrystallinity,  measurement conditions, such as the applied field direction\cite{sal16}, or sample conditions, such as the nanowire density and the concomitant inter-wire magnetostatic interactions, which are anticipated to modify the exchange biasing behavior of FM/AF core/shell nanowires. Detailed modeling in these directions is definitely highly required.
\section*{Acknowledgments}
The authors acknowledge financial support from the Special Account for Research of ASPETE through project \textit{NANOSKY} (No 80146). 
DK acknowledges the hospitality in the Department of Physics, University of Athens, where part of this work was done. 
\section*{References}

\end{document}